# Temperature-Induced Hexagonal-Orthorhombic Phase Transition in Lutetium Ferrite Nanoparticles

Olena M. Fesenko[1], Igor V. Fesych[2*], Igor V. Zatovsky[3†], Andrii D. Yaremkevych[1], Maxim Rallev[1], Andrii V. Bodnaruk[1], Eugene A. Eliseev[4] and Anna N. Morozovska[1‡]

[1]Institute of Physics of the NAS of Ukraine, Prospect Nauky, 41, Kyiv 03028, Ukraine
[2]Taras Shevchenko National University of Kyiv, Volodymyrska 64/13, Kyiv 01601, Ukraine
[3]F.D. Ovcharenko Institute of Biocolloidal Chemistry of the NAS of Ukraine, Vernadskoho Ave. 42, Kyiv 03142, Ukraine
[4] Frantsevich Institute for Problems in Materials Science, National Academy of Sciences of Ukraine Omeliana Pritsaka str., 3, Kyiv, 03142, Ukraine

## Abstract

The X-ray diffraction, Raman and infrared spectroscopies and magnetic measurements were used to explore the correlated changes of the structure, lattice dynamics and magnetic properties of the LuFeO$_3$ nanoparticles, which appear in dependence on their sintering temperature. We revealed a gradual substitution of the hexagonal phase by the orthorhombic phase in the nanoparticles, which sintering temperature increases from 700ºC to 1100ºC. The origin and stability of the hexagonal phase in the LuFeO$_3$ nanoparticles is of the special interest, because the nanoparticle in the phase can be a room-temperature multiferroic with a weak ferromagnetic and pronounced structural and ferroelectric long-range ordering. The antiferromagnetic and nonpolar orthorhombic phase is more stable in the bulk LuFeO$_3$. To define the ranges of the hexagonal phase stability, we determine the bulk and interface energy densities of different phases from the comparison of the Gibbs model with experimental results. Using effective parameters of the Gibbs model, we predict the influence of size effects and temperature on the structural and polar properties of the LuFeO$_3$ nanoparticles. Analysis of the obtained results shows that the combination of the X-ray diffraction, Raman and infrared spectroscopy, magnetic measurements and theoretical modelling of structural and polar properties allows to establish the interplay between the phase composition, lattice dynamics and multiferroic properties of the LuFeO$_3$ nanoparticles prepared in different conditions.

* Corresponding author: ihor.fesych@knu.ua
† Corresponding author: zvigo@ukr.net
‡ Corresponding author: anna.n.morozovska@gmail.com

# I. INTRODUCTION

Ferroic (ferroelectric, ferromagnetic) and multiferroic [1] nanoparticles (**NPs**) of various shape and sizes are unique model objects for fundamental studies of the surface, size, and crosstalk effects on the polar, magnetic and magnetoelectric properties [2, 3]. At the same time the NPs are very promising nanomaterials for energy storage [4, 5, 6, 7], energy harvesting and nanogenerators [8], and multi-bit memories [9, 10, 11]. Due to the surface and size effects, ultra-small (~5 − 10 nm size or less) ferroic NPs are paraelectric or/and paramagnetic in entire the temperature range. Slightly bigger magnetic ferroic NPs (~10 − 30 nm size) can be in the superparamagnetic state below the size-dependent Curie temperature. The superparamagnetic state is already used in tomography, ultra-low temperature cooling and other advanced applications of the NPs. In contrast to relatively well-studied superparamagnetic state, the superparaelectric state are not revealed yet for the ferroelectric NPs, being predicted theoretically long ago [12]. Even bigger ferroic NPs (>10 – 50 nm size) can be in the long-range ordered (anti)ferromagnetic, ferroelectric and/or multiferroelectric phases below the size-dependent Curie/Neel temperatures. In the long-range ordered phases the magnetization, $M(H)$, antiferromagnetic order, $L(H)$, and/or polarization, $P(E)$, and/or magnetoelectric responses, $M(E)$ and $P(H)$, reveal the hysteresis behavior in the electric ($E$) and/or magnetic ($H$) fields. Corresponding coercive fields, $E_c$ and $H_c$ (and thus a memory window), are controlled by the size effects. The stored information degrades very slowly in the ferroic NPs, and the storage density can be much higher for ferroelectric than for ferromagnetic NPs, however the switching times for ferromagnetic NPs can be much smaller due to the high energy barriers in ferroelectrics.

The various methods of synthesis and control of magnetic and polar properties are relatively well-developed for ferroic NPs. However, many aspects of their preparation technology still contain challenges and uncovering mysteries for fundamental theory even for the simplest case of quasi-spherical NPs of classical perovskites, such as titanates [13, 14] and orthoferrites [15, 16].

Nanosized lutetium orthoferrite, $LuFeO_3$ (**LFO**), is of great emerging interest in the field of multiferroicity. LFO has been studied by several groups in the hexagonal (**h**) phase stabilized in the nanoscale (e.g., in nanoparticles and nanofibers [8], and thin films [17]), and in the bulk orthorhombic (**o**) form [18, 19, 20, 21]. The o-LFO and h-LFO have significantly different lattice symmetry, as well as the local symmetry of the Lu and Fe cations environment [22]. This difference in the structural properties determines the dissimilarity of their physical properties.

The crystal structure of h-LFO belongs to the polar *P6₃cm* space group. The antiferromagnetic Néel temperature of h-LFO at first was determined as high as $T_{N(h)} = 440$ K. Below the Neel temperature the magnetic order in h-LFO NPs emerges from the spin reorientation resulting in a weak ferromagnetism due to the Dzhaloshinskiy-Moriya (**DM**) interaction and the single-ion anisotropy mechanism [23]. Later, $T_{N(h)}$ was reduced to 155 K and classified as the single transition to the ferromagnetically-canted antiferromagnetic state [24]. The ferroelectric Curie temperature $T_{C(h)}$ is as high as 1020 K [24], but the value of the room-temperature spontaneous polarization $P_s$ is about (4 – 6) nC/cm$^2$ [8, 16], which is more than (2 – 3) orders of magnitude lower than that for e.g., BiFeO$_3$. The spontaneous polarization expected from the polar structure of the metastable h-LFO is much higher, about 5 μC/cm$^2$ [25]. The expectation is corroborated by recent ab initio calculations [26], which predict $P_s \sim (5 - 16)$ μC/cm$^2$ below $T_{C(h)}$=1040 K due to the strong coupling between improper ferroelectricity and ferrimagnetism in the h-LFO doped with electrons. The magnetic transition near room temperature (~275–290K) and the high spontaneous magnetization, $M_s \sim (1.1 - 1.1)$ μ$_B$/Fe, are induced by a specific Fe$^{2+}$/Fe$^{3+}$ charge-ordered state [26]. Note that the morphotropic phase mixture of the h-LFO and o-LFO polymorphs, which $P_s$ may vary in a wide range, can appear in LFO thin films by suitably altering film-deposition conditions [27]. Notably, that the symmetry of the magnetic structure in the ferroelectric state of h-LFO implies that this ferroelectric can be a magnetoelectric, which ferromagnetic moment can be directly controlled by an electric field [24].

The crystal structure of the bulk o-LFO belongs to the distorted perovskite type, which is described by the nonpolar *Pbnm* space group [8, 16]. The Fe$^{3+}$ – O – Fe$^{3+}$ super-exchange interaction is very strong resulting in the high Néel temperature, $T_{N(o)} \approx (600 – 650)$ K [18]. The o-LFO phase has the largest structural anisotropy and the strongest antisymmetric spin coupling among the *R*FeO$_3$ family (*R*=Rare Earth). Due to small polar distortion of the nonpolar *Pbnm* symmetry, which probably originates from the spin canting [28], the tiny insipient spontaneous polarization (reaching 6 nC/cm$^2$ at room temperature) occurs under the second order phase transition at $T_{C(o)} \approx T_{N(o)}$ [18]. The low remanent polarization, which can be increased up to (5 - 10) μC/cm$^2$ by a special protocol of voltage application, indicate on its insipient and likely electronic nature [29]. Notably, that the magnetic and ferroelectric long-range orders coexist below 650 K and compete (being strongly coupled at the same time), because the magnetic field $H \sim 15$ kOe suppresses the remanent polarization by 95% [18]. Hence, several authors [8, 16, 18] considered the o-LFO as a multiferroic of the type-II below 650 K.

The weak ferromagnetism induced by the DM-type interactions and/or lattice frustrations is inherent to the nanoscale o-LFO below the Néel temperature. The spin frustration due to the formation of triangular lattice can induce unusual magnetic phases in the h-LFO NPs [8, 23]. Because of the strong magnetic anisotropy, the frustration in the o-LFO NPs is lower than that of the h-LFO NPs. The strong magnetic anisotropy, which support the long-range spin order, enhances the magnetic ordering and the spin reorientation temperatures in the o-LFO compared to the h-LFO NPs [8]. On the other hand, higher symmetry of the h-phase suggests a smaller entropy change ΔS between the amorphous and hexagonal phases, and thus the smaller interface energy γ. If the interface energy γ between the amorphous and hexagonal phase is smaller, ΔS can be smaller. This indicates that below a certain sintering temperature the energy barrier for the forming the h-phase can be lower than that for the forming the o-phase and suggests that the o-LFO stands on the threshold of forming the h-structure [30].

To the best of our knowledge, the influence of the preparation conditions and size effects on the structural and multiferroic properties of LFO NPs are very poorly studied. To fill the gap in the knowledge, we use the combination of the X-ray diffraction (**XRD**), Raman and Fourier-transform infrared (**FTIR**) spectroscopies, magnetic measurements and theoretical approaches to establish the influence of sintering temperature and size effects on the phase composition, lattice dynamics and multiferroic properties of the LFO NPs.

## II. MATERIALS AND METHODS

### A. Synthesis of lutetium ferrite

A single-phase $LuFeO_3$ was synthesized using sol-gel auto-combustion method by taking starting materials as lutetium nitrate $Lu(NO_3)_3 \cdot 6H_2O$ (99.99% metals basis), iron nitrate $Fe(NO_3)_3 \cdot 9H_2O$ (99.99% metals basis) and citric acid $C_6H_8O_7 \cdot H_2O$ (99.995% metals basis). This method involves exothermic and self-sustaining thermally induced redox reaction of xerogel obtained from aqueous solution, which contains metal nitrates as oxidizer and organic acid as reductant and fuel [31, 32].

According to the stoichiometric composition reactants, 0.1 mol of $Lu(NO_3)_3 \cdot 6H_2O$ and 0.1 mol of $Fe(NO_3)_3 \cdot 9H_2O$ were at first dissolved in the 100 mL of deionized water, then 0.2 mol of citric acid monohydrate was added to the solution with continuous stirring with a magnetic agitator. The molar amount of citric acid was equal to the total molar amount of metal nitrates in solution. The pH of the obtained clear mixed solution is adjusted to 7 with the addition of 25 % ammonia solution and then magnetically stirred at ambient temperature, because above pH = 6 citric acid occurs mainly in the form of $HCit^{2-}$ and $Cit^{3-}$ ions, which favors better complexing of the metal ions [33]. The homogeneous brown solution mixture was then heated at (70 – 80)°C and stirring continuously for 2 hours and evaporated to get a highly viscous and dried gel. Then the dried viscous gel is heated on a hot plate at 200°C until it gets self-ignited. Upon the ignition, the dried gel triggers burning in a self-propagating combustion reaction with rapid and vigor flames until all the gels were completely burnt out. In the combustion synthesis, besides the target product (lutetium ferrite powder), gases in the most stable form, i.e., $CO_2$, $H_2O$ and $N_2$, are produced as gel is being combusted [34, 35].

Assuming complete combustion, the schematic equation for the formation of the sample can be proposed as follows:

$Lu(NO_3)_3 + Fe(NO_3)_3 + 2C_6H_8O_7 + 1.5O_2$ (air) → $LuFeO_3 + 12CO_2 + 2N_2 + 8H_2O$.

Schematic illustration of the LFO sintering process is shown in **Fig. 1(a)**.

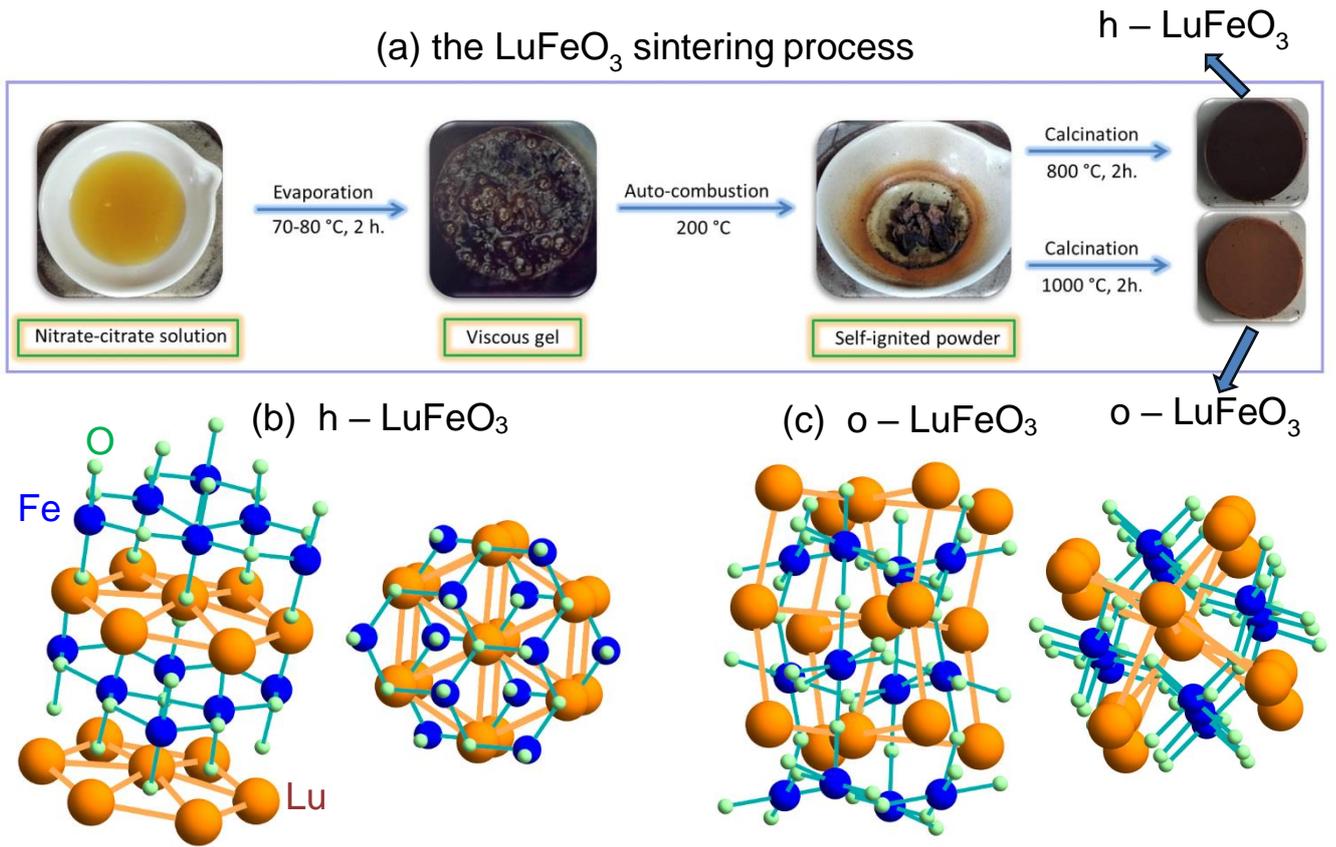

**Figure 1. (a)** Schematic illustration of LFO sintering process by a sol-gel auto-combustion method. Ball-and-stick models of hexagonal **(b)** and orthorhombic **(c)** LFO crystal lattice.

Finally, the powder was calcined at different temperatures (700°C, 800°C, 900°C, 1000°C and 1100°C) for 2 hours with heating and cooling rates of 5°C/min and obtained five samples were named as the LFO-700, LFO-800, LFO-900, LFO-1000 and LFO-1100, respectively.

**B. Phase formation in the Lu-Fe-O system according to the powder X-Ray diffraction**

Regarding the processes of phase formation in the Lu-Fe-O system, there is a lot of information that explains the reason for co-crystallization together with the h- and o-phases. The information confirms that the formation of the h-phase occurs at first (the first recalescence), given the lower formation enthalpy [36]. However, at the grain boundary, a thermodynamically more stable o-phase is formed in parallel due to the difference in the surface energy of the two phases. When the temperature increases, the metastable h-phase changes to the more stable o-phase (the second recalescence). The size effect influences the coexistence of the h- and o-phases. Moreover, the sizes of the h-phase nanocrystalline regions are always smaller than the regions of o-phase.

Crystal structure and phase analysis were characterized by means of the XRD using a LabX XRD-6000 diffractometer (Shimadzu, Japan) with Cu-Kα1 radiation. The X-Ray tube was operated at the current of 35 mA and the voltage 40 kV. The exposure time was 1 s, and the measured angle (2$\theta$) was from 5 to 70°. The scanning step was 0.02°. To identify the crystallographic phases in the studied system we used the Match software [37] and database of the International Committee for Powder Diffraction Standards (JCPDS PDF-2). All the XRD patterns were analyzed with the Fullprof Suite program (version July 2017) [38, 39] by employing the Rietveld refinement technique [40, 41].

According to the phase analysis of the samples from the XRD data, it was established that after passing through the self-combustion process of the nitrate-citrate gel (additional annealing at 600°C), crystalline phase was not formed (see **Fig. A1** in **Appendix A** in Ref.[42]). The formation of mixtures of hexagonal (space group $P6_3cm$) and orthorhombic (space group $Pbnm$) LFO phases is observed in the sintering temperature range of (700 – 900)°C. The amount of orthorhombic perovskite structure reaches a maximum and exceeds 90% at 800°C. When the sintering temperature increases to 1000°C, the h-phase of LFO completely disappears being substituted by the o-phase. Rietveld refinement of crystallographic parameters and unit cell volume for LFO nanopowders sintered at different temperatures are listed in **Table I.** The phase fractions calculated from the Rietveld analysis of the XRD pattern are shown in the **Figure 2**.

**Table I.** Rietveld refinement[*] of crystallographic parameters for LFO nanopowders sintered at different temperatures $T_S$

| Sample | LFO-700 | | LFO-800 | | LFO-900 | | LFO-1000 | LFO-1100 |
|---|---|---|---|---|---|---|---|---|
| $T_S$ [°C] | 700 | | 800 | | 900 | | 1000 | 1100 |
| Crystal system | orthorhombic | hexagonal | orthorhombic | hexagonal | orthorhombic | hexagonal | orthorhombic | orthorhombic |
| Space group | $Pbnm$ (No. 62) | $P6_3cm$ (No. 185) | $Pbnm$ (No. 62) | $P6_3cm$ (No. 185) | $Pbnm$ (No. 62) | $P6_3cm$ (No. 185) | $Pbnm$ (No. 62) | $Pbnm$ (No. 62) |
| Lattice parameters [Å] | $a$=5.2110(3) $b$=5.5361(4) $c$=7.5601(4) | $a$=5.9639(7) $b$=5.9639(7) $c$=11.7848(9) | $a$=5.2130(1) $b$=5.5486(1) $c$=7.5618(1) | $a$=5.9585(2) $b$=5.9585(2) $c$=11.7420(9) | $a$=5.2081(1) $b$=5.5439(1) $c$=7.5525(1) | $a$=5.9698(6) $b$=5.9698(6) $c$=11.7009(2) | $a$=5.2107(1) $b$=5.5489(1) $c$=7.5581(1) | $a$=5.2098(1) $b$=5.5457(1) $c$=7.5547(2) |
| Volume [Å³] | 218.101(2) | 363.019(1) | 218.729(8) | 361.034(3) | 218.067(8) | 361.141(8) | 218.536(7) | 218.277(8) |
| Size [nm]** | 53,8 ± 13,6 | 8,6 ± 0,6 | 61,4 ± 7,2 | 12,7 ± 0,7 | 95,7 ± 0,3 | 27,4± 1,8 | 103,7 ± 4,2 | 130,7 ± 3,9 |

*Background - Chebyschev polynomial, Wavelength of CuKα Radiation, $K_{\alpha}1$=1.54059, $K_{\alpha}2$=1.54431 [Å]

** "Size" is the size of the coherent scattering regions of the o-phase and h-phase, $D_o$ and $D_n$, respectively. For example, $D_o = (53,8 \pm 13,6)$ nm and $D_n = (8,6 \pm 0,6)$ nm for $T_s$ =700°C.

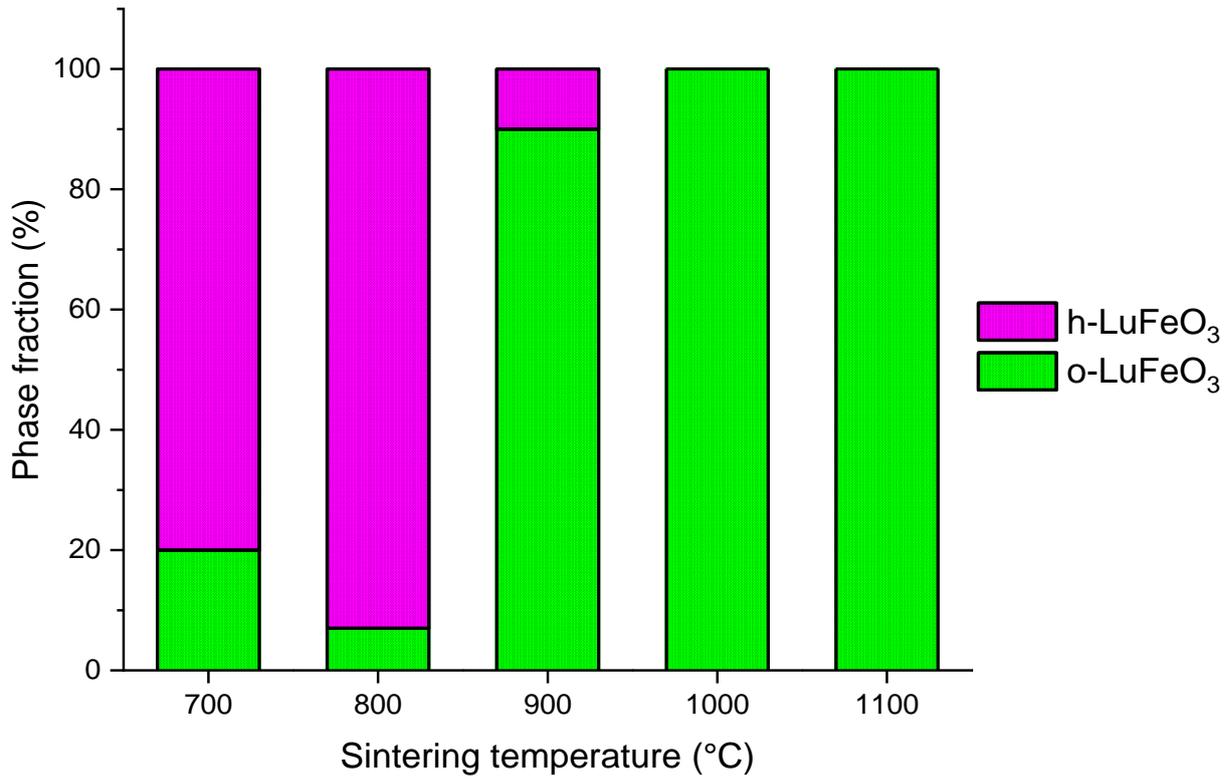

**Figure 2.** The phase fractions determined from the Rietveld analysis of the XRD pattern.

The crystallite size of the LFO nanopowders was calculated from the X-ray line broadening using the Scherrer formula:

$$D_{hkl} = \frac{K \cdot \lambda}{\beta_{hkl} \cdot \cos\theta_{hkl}} \quad (1a)$$

where $D_{hkl}$ in nm is the average crystallite size along the direction normal to the diffraction plane (*hkl*), K is the shape factor equal to 0.9, λ is the X-ray wavelength (0.15406 nm) of $Cu_{K\alpha}$-radiation, $\beta_{hkl}$ is the integral breadth of the peak related to the diffraction plane (hkl), and $\theta_{hkl}$ is the Bragg angle in radians for the crystallographic plane (hkl). The true integral peak width was calculated from the formula:

$$\beta_{hkl} = \sqrt{\beta_{exp}^2 - \beta_0^2}, \quad (1b)$$

where $\beta_{exp}$ is the experimental peak width of the sample at half maximum intensity. $\beta_0$ is the instrumental broadening of the diffraction line, which depends on the technical features of the diffractometer. The powder X-ray diffraction patterns showed that the integral breadth of the Bragg reflections is described by the pseudo-Voigt function with a large (up to 90% or more) contribution of the Lorentz function (η → 1) in the interval (5 – 30)°. Therefore, the Lorentz function was chosen to describe the shape of the diffraction peaks. To exclude the instrumental broadening $\beta_0$, a standard silicon X-ray powder diffraction data was recorded under the same condition. The (002) peak of the h-phase and (111) peak of the o-phase of LFO were chosen for calculations as the most suitable for crystallite size determination [compare **Fig. 3(a)** and **3(b)**]. Rietveld refined XRD patterns for the LFO nanopowders sintered at 800°C and 1000°C, which are the most interesting cases, are shown in **Fig. 3(c)** and **3(d)**, respectively.

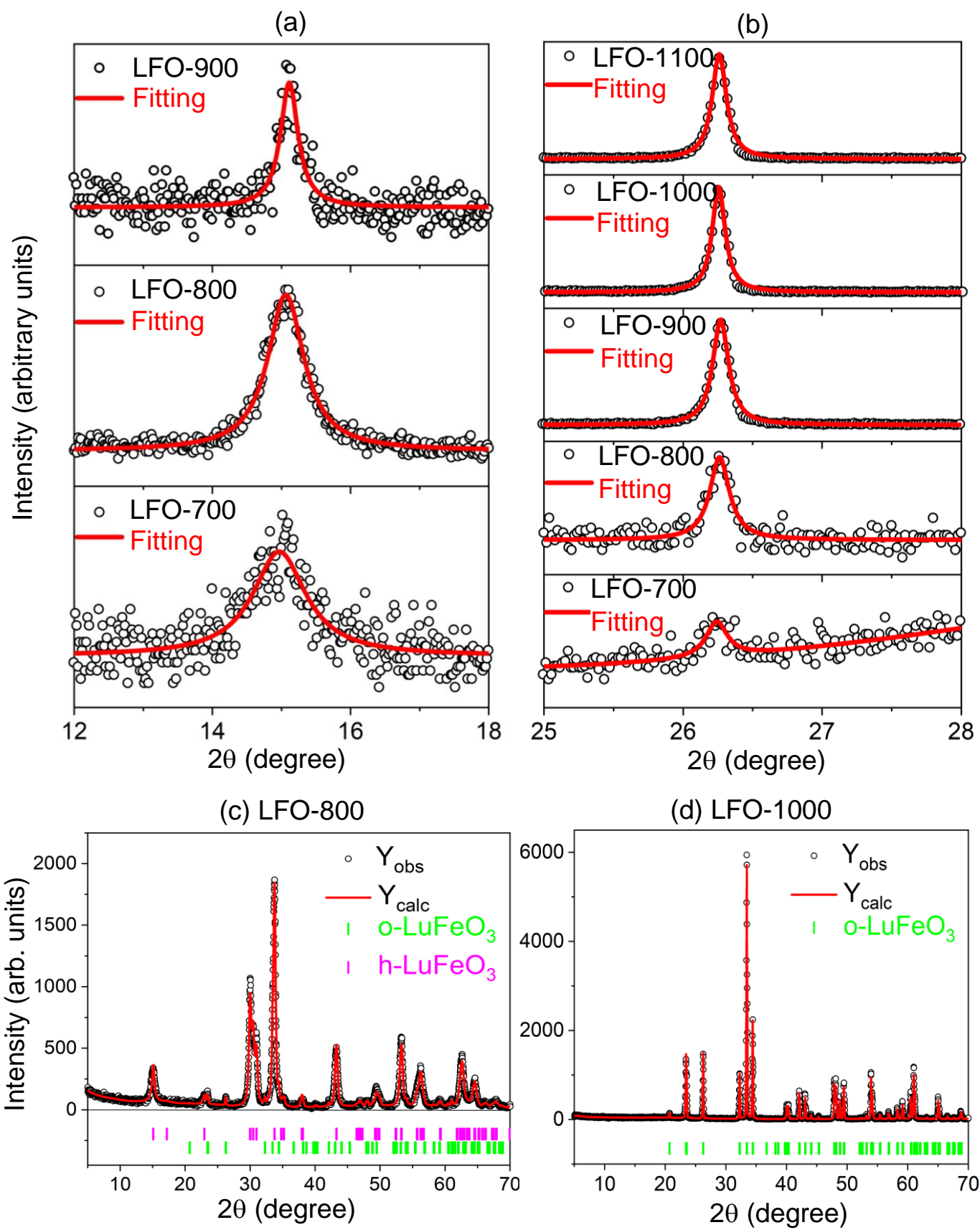

**Figure 3.** Rietveld refined XRD patterns for the LFO NPs synthesized at 700°C - 1100°C. The (002) peak of the hexagonal phase (left column, **(a)**) and the (111) peak of the orthorhombic phase (right column, **(b)**) of LFO

nanopowders: observed experimentally (empty black circles) and fitted using the Lorentz function (red solid curves). Rietveld refined XRD patterns for the LFO NPs synthesized at 800°C **(c)** and 1000°C **(d)**. The observed experimental diffractograms are shown by empty black circles ($Y_{obs}$); red curves ($Y_{calc}$) are calculated diffractograms. The vertical scale is the intensity in arbitrary units. The magenta and green vertical bars have no relation to the vertical scale; they indicate the angular positions of the allowed Bragg reflections corresponding to the hexagonal (magenta bars) and orthorhombic (green bars) phases.

Hence, the Rietveld analysis of XRD patterns reveals the gradual substitution of the h-phase by the o-phase in the LFO NPs under the sintering temperature increase from 700°C to 1100°C. Namely, we observed the following changes of the o- and h-phase fractions in dependence on the sintering temperature: 80% h-phase and 20% o-phase at 700°C, 93% h-phase and 7% o-phase at 800°C, 10% h-phase and 90% o-phase at 900°C, 100% o-phase at 1000°C and 1100°C. The average size of the h-phase coherent scattering regions increases from 8 nm to 28 nm with the sintering temperature increase. The sizes are significantly smaller than the size of the o-phase regions, which increases from 53 nm to 130 nm with the sintering temperature.

### III. MODELLING OF THE PHASE DIAGRAMS OF LuFeO$_3$ NANOPARTICLES

The energy density excess of a LFO NP, $G_{np}$, which is associated with the sizes $D_h$ and $D_o$ of the regions of coherent scattering for the h-phase and the o-phase, respectively, can be estimated within the "effective" Gibbs model [43]:

$$G_{np} = G_{np}^o + G_{np}^h + G_{int} = V_o \tilde{G}_{bulk}^o + V_h \tilde{G}_{bulk}^h + S_{int} \tilde{G}_s^{ho}, \qquad (2a)$$

Here $\tilde{G}_{bulk}^h$ and $\tilde{G}_{bulk}^o$ are the bulk energy densities of the h-phase and the o-phase, respectively. $S_{int}$ and $\tilde{G}_s^{ho}$ are the interface area and energy density, respectively.

Let us introduce the average "effective" volumes of the o- and h- phases, $V_o = \frac{\pi}{6} D_o^3 n_o$ and $V_h = \frac{\pi}{6} D_h^3 n_h$, and consider the regions as "effective" spheres with diameters $D_o$ and $D_h$, respectively. Notably that the effective volumes depend on the numbers $n_o$ and $n_h$ of "real" inclusions of the o-phases and h-phases in the NP. We would like to underline that we consider the effective spheres, while the real shape of the inclusions can be non-spherical. Also, the effective spheres may contain less or more than one particle, because the diameters $D_o$ and $D_h$ are the average sizes of the o-phase and h-phase coherent scattering regions. As a rule, the sizes of the coherent scattering regions are smaller (or even much smaller) than the physical sizes of the nanoparticles, which are shown in **Fig. 3(a)**. Since the h-phase is less stable than the o-phase in the bulk, i.e. $\tilde{G}_{bulk}^o < \tilde{G}_{bulk}^h$, we can assume that it becomes stable in the NPs due to the negative interface energy $\tilde{G}_s^{ho}$, which can be expanded in series of $\frac{1}{D_h}$. Thus, the Gibbs energy in Eq.(2a) can be re-defined as:

$$G_{np} = V_o \tilde{G}_{bulk}^o + V_h \left( \tilde{G}_{bulk}^h + \frac{\tilde{\gamma}_s^{ho}}{D_h} + \frac{\tilde{\delta}_s^{ho}}{D_h^2} + \cdots \right). \tag{2b}$$

Here we suppose that $S_{int} = \pi D_h^2 n_h$. We also assume that the interface energy density should be negative, $\frac{\gamma_s^{ho}}{D_h} + \frac{\delta_s^{ho}}{D_h^2} < 0$, at least in some range of sizes and sintering temperatures $T_S$. One can regard that the o-phase is stable for $D_o \gg D_h$, i.e. in the bulk case.

The geometry of Eq. (2b) can be imagined as "effective" spherical nuclei, where the h-phase is stable inside the spatial region filled by the o-phase. Also, we would like to underline that we consider the "effective" spheres, while the real shape of the NPs is non-spherical. Moreover, the grain boundary energy density in a single phase may also affect the particle size. Hence, the Gibbs model is oversimplified and requires significant modifications allowing for the NPs morphology. Also, we assume that the thermal equilibrium determines the phase formation. If this is not the case, the kinetics of the reaction should be investigated.

Note that the expression (2b) does not account for the possible anisotropy of the interface energy, being the simplest scalar approximation. Often, h-LFO nuclei are isomorphic and mainly two-dimensional (2D), and o-LFO nuclei are often polymorphic and three-dimensional (3D) [44]. The anisotropy, which leads to the modifications in Eq.(2), requires the proper analysis of the samples morphology.

Indeed, the SEM images of LFO NPs sintered at 800°C and 1000°C, shown in **Fig. 4(a)**, revealed that the surface morphology of the LFO-800 and LFO-1000 samples are strongly different and contains 2D and 3D features, respectively. Indeed, it is seen from the SEM images that the morphology of the LFO-800 sample, which contains 80 % of the h-phase, contains the dominant amount of plate-like (or flake-like) regions. At the same time, the morphology of the LFO-1000 sample, which contains 100 % of the o-phase, contains the dominant amount of rounded droplet-like regions. The particles are highly agglomerated independently on the annealing temperature. The formation of agglomerates covered by smaller particles is typical for different sol-gel derived compounds. Compounds, sintered at lower temperatures, also show open porous in surface microstructure. The morphology of LFO NPs, revealed by SEM, are very similar to those of rare-earth iron garnets [45].

The Gibbs model [43] is comprehensive and widely used for the calculations of the phase diagrams and related properties, but the parameters, $\tilde{G}_{bulk}^o$, $\tilde{G}_{bulk}^h$, $\tilde{\gamma}_s^{ho}$ and $\tilde{\delta}_s^{ho}$ are unknown for the LFO. Thus, at first, we should estimate the parameters from the XRD results.

A relative content of the o-phase, $x$, can be found from the well-known Gibbs formula for the statistical probability, and is given by the expression:

$$x = \frac{\exp\left[-\frac{G_{np}^o}{k_B T_p}\right]}{\exp\left[-\frac{G_{np}^o}{k_B T_p}\right]+\exp\left[-\frac{G_{np}^h}{k_B T_p}\right]} \equiv \frac{1}{1+\exp\left[\frac{\pi}{6k_B T_p}(D_o^3 \tilde{G}_{bulk}^o n_o - \{D_h^3 \tilde{G}_{bulk}^h + \tilde{\gamma}_s^{ho} D_h^2 + \tilde{\delta}_s^{ho} D_h\} n_h)\right]} \quad (3)$$

Where $T_p = T_S + \Delta$ is the sintering temperature in Kelvins, $T_S$ is the sintering temperature in °C, $\Delta = 273$ K and $k_B = 1.38 \cdot 10^{-23}$ J/K. It is possible to find the re-designated "effective" parameters,

$$G_{bulk}^o = \frac{n_o}{6}\tilde{G}_{bulk}^o, \quad G_{bulk}^h = \frac{n_h}{6}\tilde{G}_{bulk}^h, \quad \gamma_s^{ho} = \frac{n_h}{6}\tilde{\gamma}_s^{ho}, \quad \delta_s^{ho} = \frac{n_h}{6}\tilde{\delta}_s^{ho}, \quad (4a)$$

from the system of linear equations:

$$\left(D_{oi}^3 G_{bulk}^o - D_{hi}^3 G_{bulk}^h\right) - \gamma_s^{ho} D_{hi}^2 - \delta_s^{ho} D_h = \frac{k_B}{\pi}(T_{Si} + \Delta)\log\left(\frac{1}{x_i} - 1\right). \quad (4b)$$

Here $\{x_i, D_{hi}, D_{oi}\}$ are the experimental data sets (see **Table I**), corresponding to different sintering temperatures $T_{Si}$, and $i = 1, 2, 3, \ldots$. In result of the fitting, we obtained that:

$$G_{bulk}^o = -116.1 \text{ J/m}^3, \quad G_{bulk}^h = 6.5492 \cdot 10^4 \text{ J/m}^3, \quad (5a)$$

$$\gamma_s^{ho} = -2.05697 \cdot 10^{-3} \text{ J/m}^2, \quad \delta_s^{ho} = 6.855 \cdot 10^{-12} \text{ J/m}, \quad (5b)$$

$$D_h(T_S) = 276.7 - 0.754 T_S + 0.00053 T_S^2 \text{ (nm)}, \quad (5c)$$

$$D_o(T_S) \cong -16.89 + 0.03539 T_S + 0.00008929 T_S^2 \text{ (nm)}. \quad (5d)$$

As anticipated, $G_{bulk}^o < 0$ and $G_{bulk}^o \ll G_{bulk}^h$, and therefore the h-phase can be stable in the NPs due to the gain in the interface energy, $G_s^{ho} = \frac{\gamma_s^{ho}}{D_h} + \frac{\delta_s^{ho}}{D_h^2}$, which is negative for the sizes $D_h > -\frac{\delta_s^{ho}}{\gamma_s^{ho}}$. The critical size, $D_{cr}$, where both phases have equal energies, satisfies the condition $G_{np}^h(D_{cr}) = G_{np}^o$. For the parameters given by Eqs.(5), $D_{cr} \approx 27.5$ nm. For $D < D_{cr}$ both phases can coexist in the LFO NPs, and the h-phase has smaller energy. For $D > D_{cr}$ the o-phase is stable, and the h-phase can be either metastable or unstable in comparison with the o-phase. The sizes of h-phase metastability correspond to the condition $G_{np}^h < 0$, which gives $D_{m1} < D < D_{m2}$, where $D_{m1} \approx 1$ nm and $D_{m2} \approx 30$ nm are the roots of the quadratic equation $G_{bulk}^h + \frac{\gamma_s^{ho}}{D_h} + \frac{\delta_s^{ho}}{D_h^2} = 0$.

The dependence of the o-phase and h-phase fractions, $x$ and $1-x$, on the sintering temperature $T_S$ are shown in **Fig. 4(b)** by red and blue curves with symbols, respectively. Symbols correspond to the data extracted from XRD spectra. Solid curves are approximation functions given by Eq.(3) with the parameters $G_{bulk}^o$, $G_{bulk}^h$, $\gamma_s^{ho}$ and $\delta_s^{ho}$ given by Eqs.(5).

The dependences of the average sizes $D_h$ and $D_o$ on the sintering temperature $T_S$ are shown in **Fig. 4(c)** by red and blue curves with symbols, respectively. Symbols correspond to the data extracted from the XRD spectra; solid curves are approximation functions given by dependences (5c) and (5d).

The average size of the h-phase nanocrystalline regions increases from 8 nm to 28 nm with $T_S$ increase, being significantly smaller than the size of the o-phase region, which increases from 53 nm to 130 nm with $T_S$ increase.

The dependence of the h-phase (blue curve + symbols) and o-phase (red line) energy densities on the average size $D$ of the regions of coherent scattering are shown in **Fig. 4(d).** The red line is straight because it is the constant bulk energy of the o-phase, $G_{bulk}^o$. It is seen that the solid blue curve, which is the approximation given by Eq. (2) with the parameters given by Eq.(5) for nonzero $\gamma_s^{ho}$ and nonzero $\delta_s^{ho}$, describes the XRD data much better than the dashed curve, which is the approximation function for nonzero $\gamma_s^{ho}$ and zero $\delta_s^{ho}$.

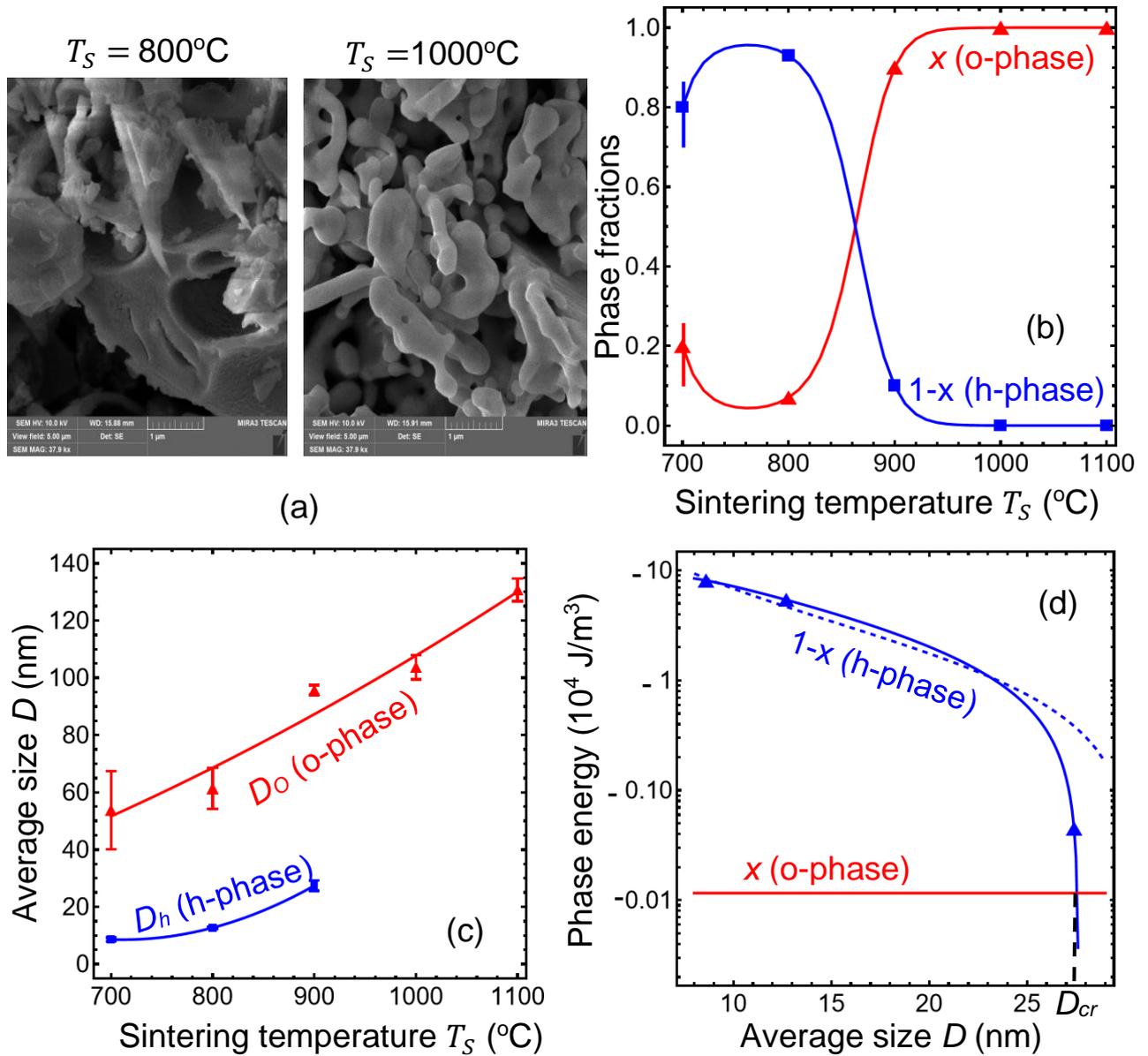

**Figure 4. (a)** SEM images of LFO nanostructures sintered at 800°C (left image) and 1000°C (right image). **(b)** The dependence of the o-phase (red curve + symbols) and h-phase (blue curve + symbols) fractions, $x$ and 1-$x$,

on the sintering temperature $T_S$. **(c)** The average sizes $D_h$ (blue curve + symbols) and $D_o$ (red curve + symbols) of the coherent scattering regions in dependence on $T_S$. **(d)** The dependence of the h-phase (blue curve + symbols) and the o-phase (red line) energy densities on the average size $D$ of the coherent scattering regions. Symbols correspond to the data extracted from XRD spectra. The solid curve is the approximation function given by Eq. (2) with the parameters given by Eq.(5). The dashed curve is the approximation function given by Eq. (2) for nonzero $\gamma_S^{ho}$ and zero $\delta_S^{ho}$.

## IV. RAMAN AND FTIR SPECTRA OF LuFeO$_3$ NANOPOWDER

According to the XRD data (see the **Fig. 2** and **3**), it was found that the almost pure h-phase is formed after sintering at 800°C. The h-phase transforms in the o-phase at higher temperatures. For sintering at $T_S = 1000$°C and higher the pure o-phase is thermodynamically stable and no further structural transitions happen with increasing the sintering temperature. Therefore, the two values $T_S = 800$°C and $T_S = 1000$°C, as the most representative cases, are chosen to study all the main spectral, magnetic and polar properties in dependence on $T_S$. The influence of the phase inhomogeneity and the degree of crystallinity on the physicochemical parameters of the LFO NPs in the vicinity of expected morphotropic region of the o-h transition (e.g., for $750°C < T_S < 800°C$) will be studied elsewhere.

Raman spectra were measured using a Renishaw InVia (England) micro-Raman spectrometer equipped with a DM2500 Leica confocal optical microscope. A laser operating at a wavelength of λ = 633 nm was used to measure the Raman scattering spectra of LFO NPs. Processing of Raman spectra was performed using the WiRE 5.2 program, which was used to determine the peaks and decompose the bands into components. All measurements were performed at room temperature. The FTIR spectra were measured using a Vertex 70 FTIR spectrometer, Bruker (Germany).

Typical Raman spectra of LFO nanopowders are shown in **Fig. 5**. For the LFO-800 sample sintered at 800°C, peaks at 264 cm$^{-1}$, 404 cm$^{-1}$ correspond to A1, E1 Lu-O vibrations, which were observed and described in Refs. [16, 46]. The peak at 475 cm$^{-1}$ corresponds to bending vibrations of Fe-O$_6$, and the peak at 645 cm$^{-1}$ corresponds to Fe-O$_6$ stretching vibrations. The conclusion follows from the analysis of the Raman spectra by the symmetry theory and is in a full agreement with the known Raman spectra of LFO NPs (see e.g., right columns in Table 1 in Ref. [16], where corresponding peaks are located at 477 cm$^{-1}$ and 644 cm$^{-1}$, respectively). All these peaks, except for 645 cm$^{-1}$, are typical for the h-phase of LFO, whereas the stretching vibrations of Fe-O$_6$ should be at 651 cm$^{-1}$ (see e.g., the middle columns in Table 1 in Ref. [16]). However, the presence of this peak can be explained by Fe-O$_5$ bipyramidal tilt, which potentially causes the shift in the Raman modes in NPs and thin films [16].

In the case of the LFO-1000 sample sintered at 1000°C, the peak at 240 cm$^{-1}$ corresponds to the rotational vibrations of Fe-O$_6$ octahedra, peaks at 350 cm$^{-1}$ and 419 cm$^{-1}$ correspond to the Lu-O vibrations, and the broad peak with maximum at 650 cm$^{-1}$ corresponds to Fe-O$_6$ octahedra stretching vibrations [16]. This indicates the presence of both the o-phase and the h-phase in the LFO-1000 sample [16, 47]. However, the XRD data, shown in **Fig. 3(c)**, reveals only the o-phase in the sample. However, in accordance with theoretical modelling, which results are shown in **Fig. 4**, the metastable h-phase can exist in the LFO NPs sintered at 700°C< $T_S$ <1050°C. Note that the Raman peak at ~650 cm$^{-1}$ corresponds to the A$_1$ mode related with the ferroelectric phonon mode that softens at much higher ferroelectric Curie temperature $T_C$ ~ 1020 K [24].

Possible reason of the discrepancy between the analysis of the XRD and Raman spectra is the surface effect. We can assume this, because Raman signal can be more sensitive to the surface layer of the material in comparison with its bulk. The increased sensibility comes from several reasons, such as local surface fields, electric charges redistribution and/or appearance of surface modes, localized in the thin shell covering the NP core. In result the relative fraction of the Raman-active modes corresponding to the shell can be significantly higher than the core fraction in the recorded Raman spectra. The XRD spectra is equally sensitive to the surface and bulk of the NP, and thus the XRD information about relative fraction of different phases is indifferent on their location. Since the surface energy coefficient $\gamma_s^{ho}$ in Eq.(2b) appears negative (from the best fitting of XRD data), we can conclude that the sample surface should contain bigger (or even much bigger) fraction of the h-phase regions. At the same time the core of the NPs can be in the o-phase or in the h-phase, in dependence on the sintering temperature $T_S$ and other factors. Due to the gain in the interface energy, the h-phase can be stable in the thin shell of NPs, which thickness decreases with $T_S$ increase, meanwhile the size of the o-phase core increases. In result, it may happen that the XRD cannot resolve the weak signal from the ultra-thin shell at $T_S$> 900°C because its volume fraction becomes less than the noise/signal threshold. At the same time the Raman signal from the shell can be resolvable still.

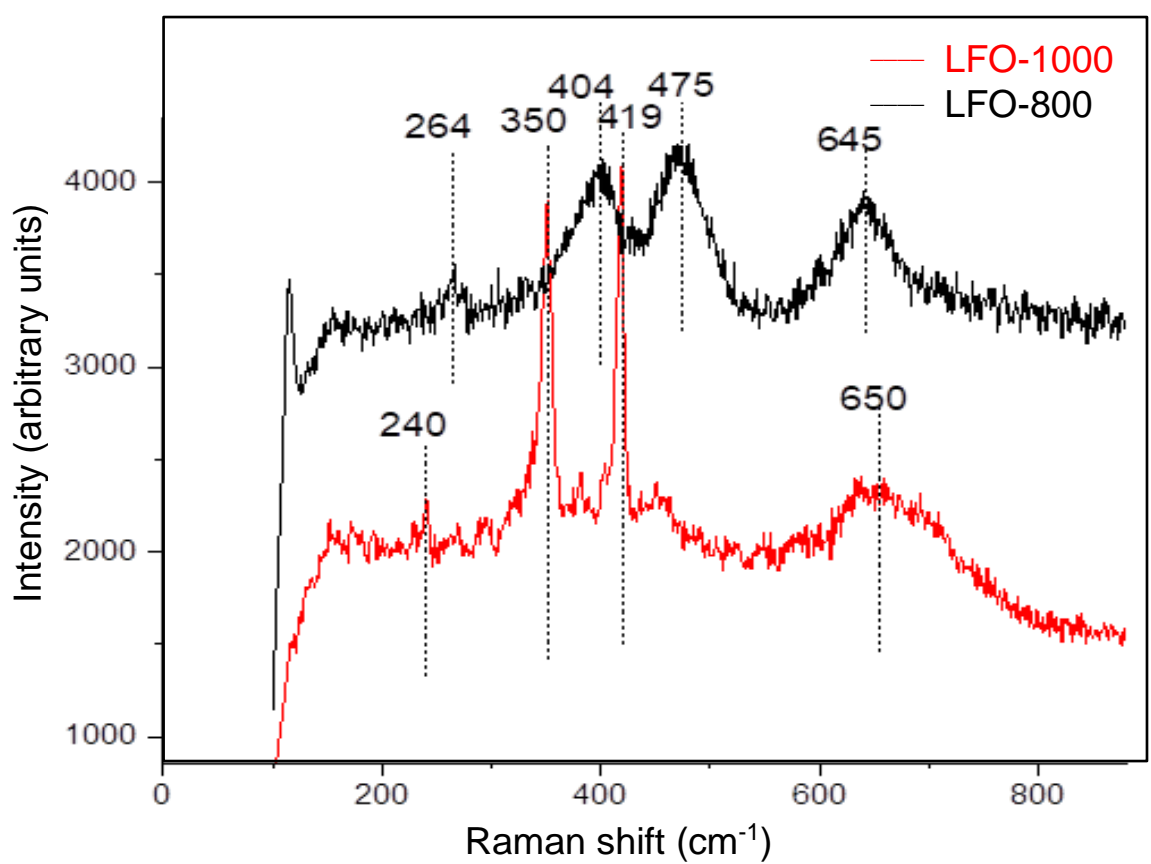

**Figure 5.** Raman spectra of LFO nanopowders sintered at 800°C (black spectrum, sample LFO-800) and 1000°C (red spectrum, sample LFO-1000). Vertical dotted lines correspond to the spectral peaks and the numbers above the vertical lines correspond to the maxima of peaks in cm$^{-1}$, which positions are determined instrumentally by the special software.

Typical FTIR spectrum of LFO-800 and LFO-1000 nanopowders in the KBr matrix in the region 400 – 900 cm$^{-1}$ is shown in **Fig. 6**. According to [48], only two major bands in the frequency range of 250 – 600 cm$^{-1}$ can be observed for rare-earth ferrites, which are associated to the Fe-O stretching. For the LFO-1000 sample, these two bands are located at 546 and 440 cm$^{-1}$ (see **Fig. 6**), and the FTIR spectra is like those reported earlier for LuFeO$_3$ [49]. For the LFO-800 sample, only one band at 468 cm$^{-1}$ is clearly observed, and the slope of the FTIR curve indicates the presence of another band with a frequency number below 400 cm$^{-1}$. As is well known, the wave number of the IR bands can be influenced by the symmetry of the crystal lattice, the coordination number, and the bond length [50]. This implies that for the h-LFO the center of the second band is below the value of 400 cm$^{-1}$. In addition, in the FTIR spectrum of LFO-800 there are weak absorption bands around the value of 546 cm$^{-1}$ (see the red curve in **Fig. 6**), which may indicate the presence of the o-LFO and correlate well with the XRD results (see **Fig. 2**).

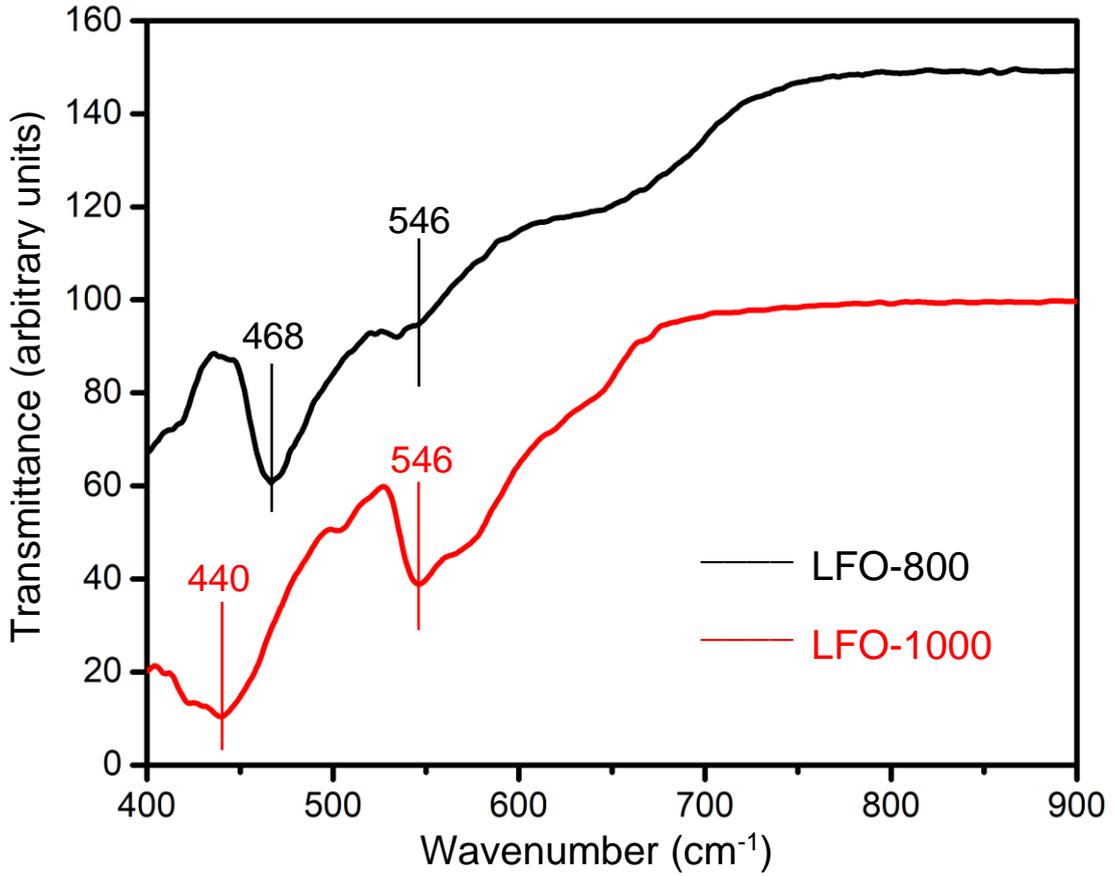

**Figure 6.** FTIR spectra of LFO nanopowders sintered at 800°C (black curve, sample LFO-800) and 1000°C (red curve, sample LFO-1000) in the region 400–900 cm$^{-1}$. Thin black lines correspond to the FTIR spectra minima and the numbers above the lines correspond to the minima in cm$^{-1}$.

## V. MAGNETIC, STRUCTURAL AND POLAR PROPERTIES

### A. Magnetization measurements

To measure the magnetization, we used a vibrating magnetometer (LDJ-9500, LDJ Electronics, Troy, MI 48099, USA) with a maximal magnetic field of 10 kOe and a thermal stabilization system with purging with vaporous liquid nitrogen. Temperature measurements of magnetization were carried out in the range of 100 K – 350 K. The samples were packed in polypropylene tubes with a diameter of 2.5 mm and a height of 8 mm.

The temperature dependence of LFO NPs magnetization $M(T)$ was measured in the magnetic field $H = 10$ kOe. Typical results are shown in **Fig. 7.** For the h-phase sample LFO-800 the magnetization quasi-linearly and very weakly increases with the increase of inverse temperature. Namely, $M(T) \sim \frac{1}{T} + M_0$, as it can be seen from the comparison of the measured magnetization curve (black symbols) and black double dot-dashed line in **Fig. 7(a)**. Here $M_0$ is negligibly small and can be related with an experimental error. The behavior $M(T) \sim \frac{1}{T}$ is characteristic for the paramagnetic phase.

The relatively small features on the magnetization curve (marked by black vertical arrows) can be associated with the appearance of a surface-induced weak ferromagnetic state in the temperature range between $T \cong 169$ K and $T \cong 312$ K. Magnetization reversal curves, $M(H)$, measured for the LFO-800 sample at 100 K and 350 K, are shown in **Fig. 7(b)** by the dashed and solid black curves, respectively. These curves are hysteresis-less and have a paramagnetic form.

For the LFO-1000 sample, a nonlinear strong decrease of magnetization under the temperature increase is observed, which is a characteristic feature of the superparamagnetic or weak ferromagnetic states. The cross-section of the linear approximations for magnetization measured at lower temperatures (shown by the dot-dashed red line) and higher temperatures (shown by the dotted red line) corresponds to the value $T \cong 200$ K. This value is higher than the Neel temperature of a bulk o-LFO ($T_N = 155$ K), indicating that the weak canted ferromagnetism does exist below the Neel temperature. Since the tilt angles of the approximations for magnetization (i.e., the tilt of dot-dashed and dotted red lines) with respect to the 1/T axis is not small, and magnetization values are significantly higher than the corresponding values for the paramagnetic LFO-800 NPs. Hence, the superparamagnetism may exist in the LFO-1000 NPs up to the high temperatures (due to the strong magnetoelectric coupling). The behavior may be like that in the bulk o-LFO, where the magnetic and ferroelectric long-range orders coexist below 650 K [18]. The hysteresis-less $M(H)$ curves, measured for LFO-1000 at 100 K and 350 K, are shown in **Fig. 7(b)** by the dashed and solid red curves, respectively. These hysteresis-less curve for 100 K has a pronounced superparamagnetic form, and the $M(H)$ curve for 350 K has a less pronounced, but still nonlinear $H$-dependence. Hence the magnetic measurements reveal the features of the paramagnetism in the LFO-800 samples and a mixture of weak ferromagnetism and strong (or weak) super-paramagnetism in the LFO-1000 samples.

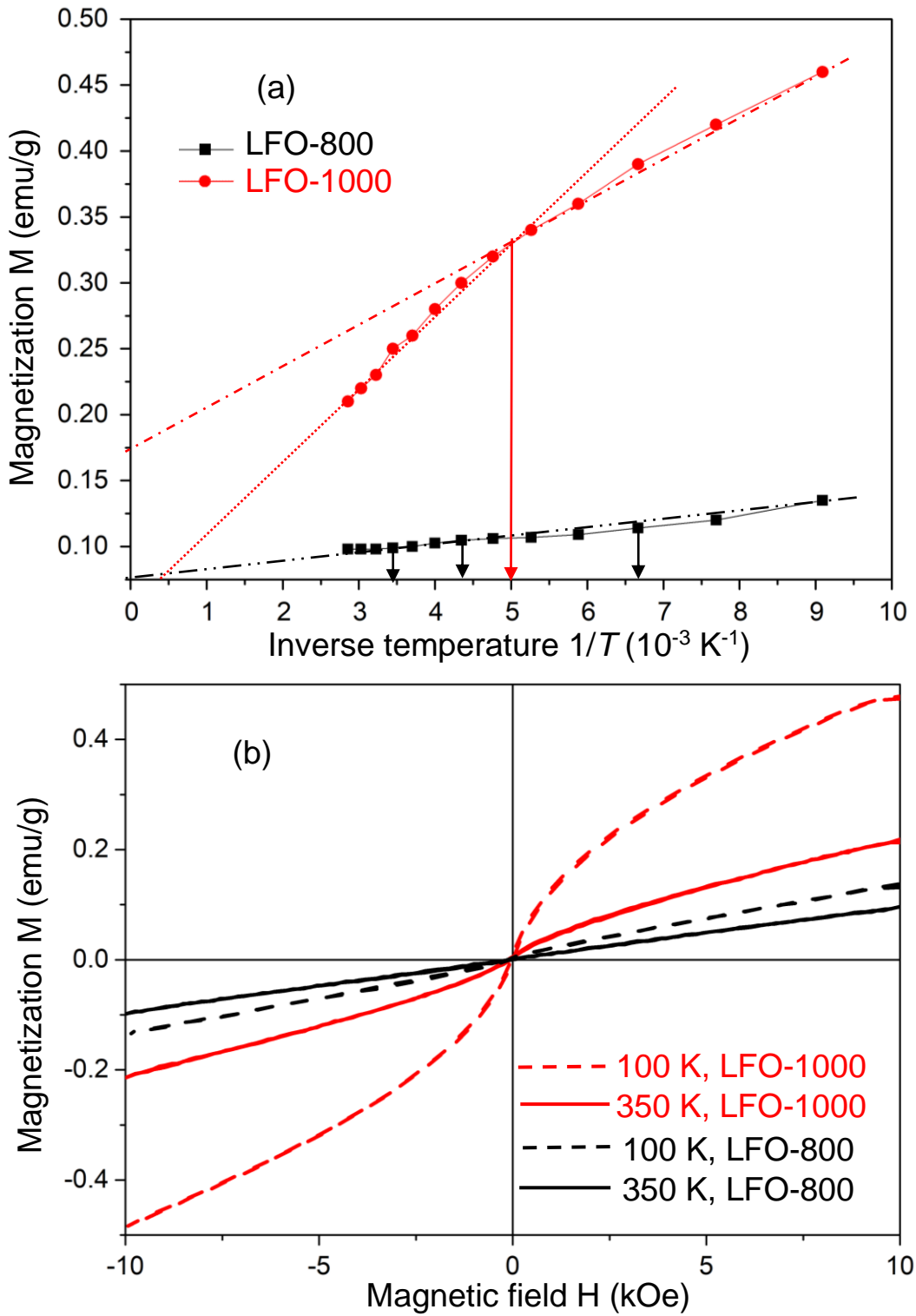

**Figure 7. (a)** Temperature dependence of the LFO NPs magnetization after sintering at 800°C (black symbols, the sample LFO-800) and 1000°C (red symbols, the sample LFO-1000). The magnetic field $H = 10$ kOe. Red dotted and dot-dashed lines are linear approximations for the magnetization of the LFO-1000 sample valid in different ranges of inverse temperature. The black double dot-dashed line is a linear approximation for the magnetization of the LFO-800 sample. **(b)** Magnetization reversal curve, $M(H)$, measured for the LFO-800 (dashed and solid black curves) and LFO-1000 (dashed and solid red curves) samples at 100 K and 350 K, respectively.

### B. Structural and polar properties calculations

Experimental measurements of a nanopowder polar properties are complicated and very often impossible without serious changes of the NPs state (e.g., by making the dense ceramics from single particles). Because of this it seems reasonable to calculate the polar properties of LFO NPs using the information about the role of surface and interface energy contributions estimated from XRD data. As a matter of fact, the surface and interface energy are responsible for the appearance of surface tension. The tension can be responsible for the size-induced polar and structural phase transitions in $R$FeO$_3$ NPs (see e.g., Ref.[51] and refs. therein).

We calculated structural and polar properties of LFO NPs in the h-phase using the Landau-Ginsburg-Devonshire model [52] for the coupled long-range orders in nanoscale hexagonal ferrites. The model considers the primary structural long-range order parameter, the vector $\boldsymbol{Q}$ of the apical oxygen ions displacement, and the secondary polar order parameter, the polarization vector $\boldsymbol{P}$. The appearance of $\boldsymbol{Q}$ is the structural phase transition accompanied by the tripling of the unit cell (so-called "trimerization"). Analytical calculations, which details are listed in **Appendix B** [42], lead to the coupled equations for the $\boldsymbol{Q}$ and $\boldsymbol{P}$ magnitudes:

$$\alpha_Q Q + b_Q Q^3 + \left(c_2 - \frac{3g^2}{a_P + \mu Q^2}\right) Q^5 + \mu Q \left(\frac{gQ^3}{a_P + \mu Q^2}\right)^2 = 0, \qquad (6a)$$

$$P = \pm \frac{gQ^3}{a_P + \mu Q^2}. \qquad (6b)$$

The temperature-dependent expansion coefficients $\alpha_Q$ and $\alpha_P$ are renormalized due to the surface tension and striction effects as

$$\alpha_Q(T, D) = a_Q(T) + R_Q \left(\frac{\gamma_s^{ho}}{D} + \frac{\delta_s^{ho}}{D^2}\right), \qquad (7a)$$

$$\alpha_P(T, D) = a_P(T) + R_P \left(\frac{\gamma_s^{ho}}{D} + \frac{\delta_s^{ho}}{D^2}\right). \qquad (7b)$$

Here $D$ is the size of h-phase nanoregion. Electrostriction and trimerization-associated distortion coefficients, $R_P$ and $R_Q$, can be estimated according to the structural data in $R$FeO$_3$ [51]. Since the ferroelectricity in h-LFO is induced by the trimerization order parameter $Q$, the coefficient $a_P(T)$ is big and positive in Eq.(7b). The coefficient $a_Q$ is negative for $T < T_C$ and positive for $T > T_C$, because $a_Q(T) = a_{Q0}(T - T_C)$, where $T_C \cong 1050$ K. Therefore, electrostriction coefficient value, $R_P$, does not matter much in Eq.(7b). The coupling strength of the strain with the trimerization order parameter, $R_Q$, can be important in Eq.(7a). All material parameters included in Eqs.(6)-(7) are described in **Appendix B** and their numerical values are listed in **Table BI** [42]. The interface energy parameters $\gamma_s^{ho}$ and $\delta_s^{ho}$ were determined from the XRD experiments and listed in Eqs.(5b).

Temperature dependences of the order parameters $Q$ and $P$ are shown in **Fig. 8(a)** and **8(b)**, respectively. The results shown in **Fig. 8** are valid only for the h-phase of LFO, where the structural phase transition is accompanied by the tripling of the unit cell (the trimerization leading to the appearance of $Q \sim 1$ Å). The tripling is also accompanied by the simultaneous improper phase transition to the ferroelectric phase with a spontaneous polarization P ~ (1 – 5) μC/cm² in dependence on the sign and magnitude of $R_Q$. Positive $R_Q$ increases $Q$ and $P$ values, and the transition $T_C$; $R_Q = 0$ corresponds to the bulk hexagonal LFO; and negative $R_Q$ decreases $Q$, $P$ and $T_C$ (compare the blue, black and red curves in **Fig. 8(b)**). An increase or decrease in the polarization of LFO NPs is associated with the change in the temperature of the structural phase transition under the influence of the surface tension. The surface tension depends on the size of the h-phase nanoregions (the physical effect is like an external pressure). Thus, the interface energy parameters, $\gamma_s^{ho}$ and $\delta_s^{ho}$, and the dependence of the interface energy on the size of coherent scattering regions, determined from the XRD data and given by Eqs.(5), allow to predict the size-induced changes in the polar and structural properties of the LFO NPs. However, to quantify our predictions it is necessary to determine the sign and value of $R_Q$ from independent experiments or/and from ab initio calculations. We vary the value of $R_Q$ in a reasonable range in **Fig. 8** and can conclude that its influence is noticeable for 10-nm NPs.

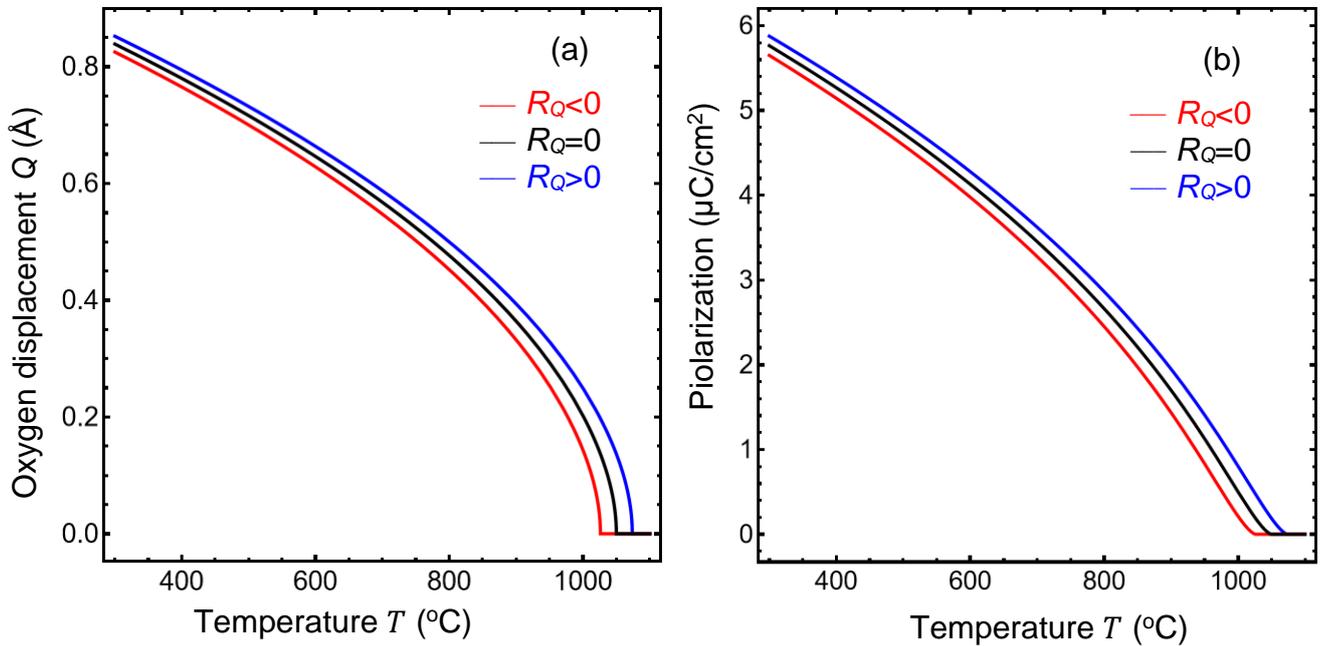

**Figure 8.** Temperature dependences of the apical oxygen ions displacement $Q$ **(a)** and spontaneous polarization $P$ **(b)** calculated for the 10-nm size of h-phase grains. Black curves are calculated for $R_Q = 0$ (which also corresponds to the bulk hexagonal LFO). Blue and red curves correspond to $R_Q = +95$ Å$^{-2}$ and $R_Q = -95$ Å$^{-2}$, respectively. Other parameters of LFO are listed in **Table BI** in **Appendix B** in Ref. [42].

Size and temperature dependences of the apical oxygen ions displacement $Q$ and spontaneous polarization $P$ calculated for LFO NPs with $R_Q < 0$ and $R_Q > 0$ are shown in **Figs. 9(a,c)** and **9(b,d)**, respectively. The contour maps look very different for $R_Q < 0$ and $R_Q > 0$. Namely, negative $R_Q$ enhances the structural and polar long-range orders for very small sizes $D < 2$ nm, and positive $R_Q$ enhances both orders for bigger sizes $D > 4$ nm. Both order parameters, $Q$ and $P$, appear and disappear simultaneously in result of the trigger-type trimerization phase transition. The size-dependent phase transition temperature, $T_{CQP}(D)$, can be found from the condition $\alpha_Q(T,D) = 0$ and is given by expression:

$$T_{CQP}(D) = T_C - \frac{R_Q}{a_{Q0}} \left( \frac{\gamma_s^{ho}}{D} + \frac{\delta_s^{ho}}{D^2} \right). \tag{8}$$

From Eq.(8), $T_{CQP} > T_C$ for $\frac{R_Q}{a_{Q0}} \left( \frac{\gamma_s^{ho}}{D} + \frac{\delta_s^{ho}}{D^2} \right) > 0$ and $T_{CQP} < T_C$ in the opposite case. Since $\gamma_s^{ho} < 0$ and $\delta_s^{ho} > 0$ in accordance with Eq.(5b), the size-induced phase transition from the multiferroic h-phase to the nonpolar phase (without any long-range order) is determined by the interplay of the two terms, $\frac{\gamma_s^{ho}}{D}$ and $\frac{\delta_s^{ho}}{D^2}$.

Hence, the values of the bulk and interface energy densities, determined from the XRD data, allow us to predict the influence of temperature and size effects on the structural and polar properties of the LuFeO$_3$ NPs in the h-phase. In accordance with our theoretical modelling, the multiferroic h-phase with a relatively big polarization ($> 5$ μC/cm$^2$) can exist in the LFO NPs at room temperature if their size varies in the range (2 – 100) nm.

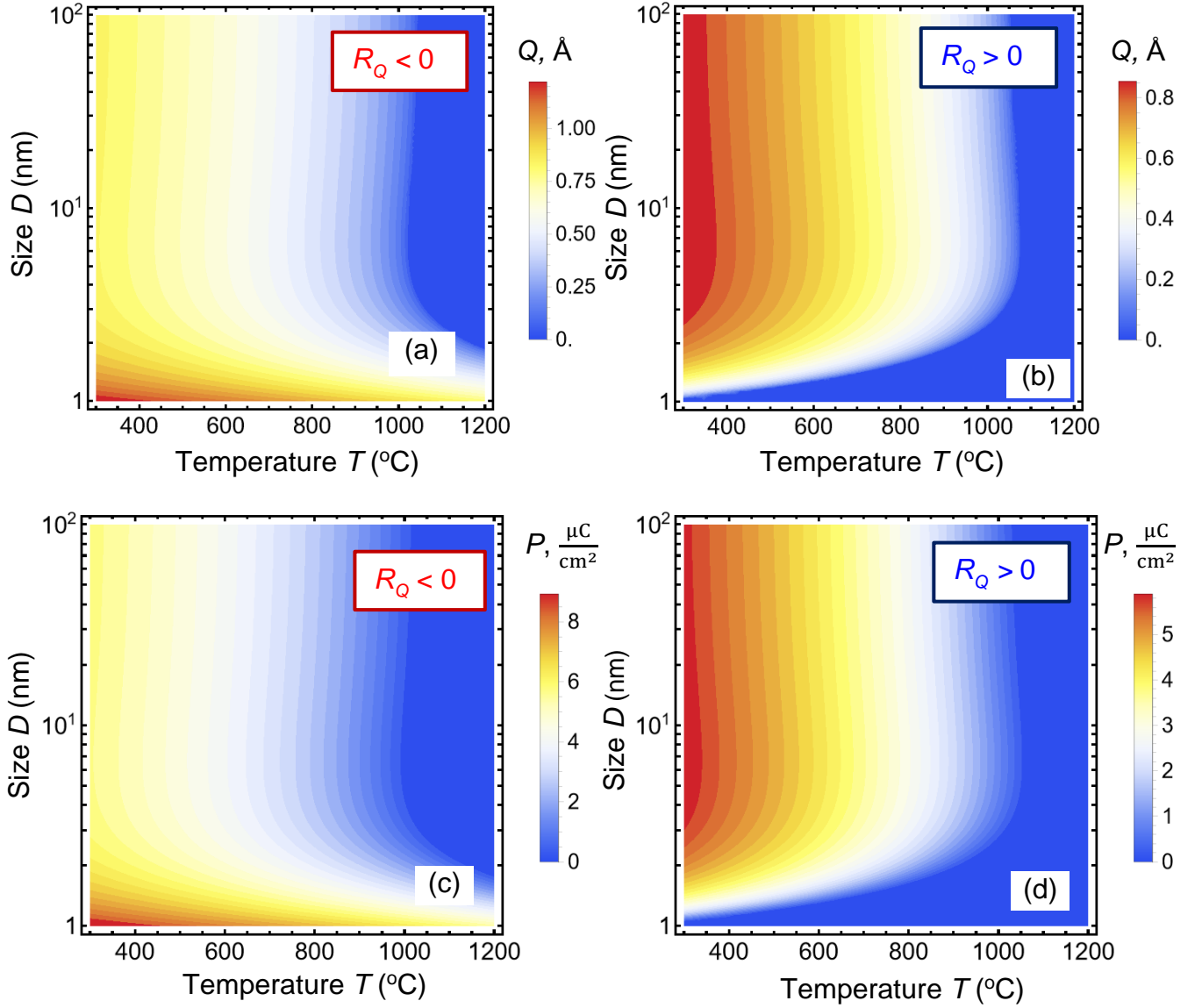

**Figure 9.** Temperature and size dependences of the apical oxygen ions displacement $Q$ (**a, b**) and spontaneous polarization $P$ (**c, d**) calculated for LFO NPs with $R_Q = -95$ Å$^{-2}$ (**a, c**) and $R_Q = +95$ Å$^{-2}$ (**c, d**) respectively. Other parameters are listed in **Table BI** in **Appendix B** in Ref. [42]. Color scales show the $Q$ and $P$ values in Å and μC/cm², respectively.

## VII. CONCLUSIONS

Rietveld analysis of XRD patterns reveals the gradual substitution of the h-phase by the o-phase in the LFO NPs under the sintering temperature increase from 700ºC to 1100ºC. To define the ranges of the h-phase stability in LFO NPs, we determine the bulk and interface energy densities of different phases from the comparison of the effective Gibbs model with experimental results. As anticipated, the bulk energy of the o-phase is big and negative, and the bulk energy of the h-phase is small and positive, and thus the h-phase becomes stable in the LFO NPs due to the gain in the negative interface energy.

The critical size, corresponding to the equal energies of the h- and o-phases, was estimated as $D_{cr} \approx$ 27.5 nm for LFO NPs sintered at (700 – 1050)°C. For the sizes $D < D_{cr}$ both phases can coexist in the LFO NPs, and the h-phase has smaller energy. For the sizes $D > D_{cr}$ the o-phase is stable, and the h-phase can be either metastable or unstable in comparison with the o-phase.

The values of the bulk and interface energy densities, determined from the XRD data, allowed to predict the influence of temperature and grain size effects on the structural and polar properties of the LFO NPs in the h-phase. In accordance with our theoretical modelling, the multiferroic h-phase can be stable in the LFO NPs if their size varies in the range (2 – 100) nm.

Raman spectra of the LFO nanopowder sintered at 800°C has the peaks typical for the h-phase. The Raman spectra of the LFO nanopowder sintered at 1000°C indicates the simultaneous presence of the o- and h-phases, while XRD data reveals only the o-phase in the sample. Possible reasons of the discrepancy between the XRD and Raman spectra are surface effects.

The temperature dependence of magnetization of LFO NPs sintered at 800°C is characteristic for the paramagnetic phase or weak ferromagnetic state in the temperature range (100 – 350) K. For the LFO NPs sintered at 1000°C the temperature dependence of magnetization is characteristic for the pronounced superparamagnetic state in the same temperature range.

In general, in accordance with XRD, Raman and FTIR data, and theoretical modelling, the sintering temperature increase from 800°C to 1000°C results in the gradual transition from the hexagonal to the orthorhombic structure in the LFO NPs. The combination of the XRD, Raman spectroscopy, magnetic measurements and theoretical approaches allows to predict the influence of size effects on multiferroic properties of the LFO NPs.

**Authors' contribution.** The research idea belongs to A.N.M. and O.M.F. I.V.F. and I.V.Z. prepared the samples and performed XRD and SEM measurements. O.M.F. and A.D.Y. performed Raman measurements. O.M.F., M.R. and I.V.Z. performed and interpreted FTIR measurements. A.V.B performed magnetic measurements. A.N.M. formulated the theoretical problem, performed analytical calculations, and compare with experiments. E.A.E. wrote the codes and performed calculations of polar and structural properties. All co-authors analyzed the obtained results and wrote corresponding parts of the manuscript. Corresponding authors worked on the manuscript improvement.

**Acknowledgments.** Authors are grateful to Referees for very useful and constructive discussions. The work (O.M.F., A.V.B., A.D.Y., M.R., and A.N.M.) is supported by the Ministry of Science and Education of Ukraine (grant № РН/ 23 - 2023, "Influence of size effects on the electrophysical properties of graphene-ferroelectric nanostructures") at the expense of the external aid instrument of the European Union for the fulfillment of Ukraine's obligations in the Framework Program of the European Union for scientific research and innovation "Horizon 2020". The work (I.V.F.) is supported by the European Union within the Project 101120397 - APPROACH. The work (E.A.E.) is supported by the DOE Software Project on "Computational Mesoscale Science and Open Software for Quantum Materials", under Award Number DE-SC0020145 as part of the Computational Materials Sciences Program of US Department of Energy, Office of Science, Basic Energy Sciences.

[51] A. N. Morozovska, E. A. Eliseev, M. D. Glinchuk, O. M. Fesenko, V. V. Shvartsman, V. Gopalan, M. V. Silibin, D. V. Karpinsky. Rotomagnetic coupling in fine-grained multiferroic $BiFeO_3$: theory and experiment. Phys.Rev. **B 97**, 134115 (2018) https://link.aps.org/doi/10.1103/PhysRevB.97.134115

[52] S. Artyukhin, K. T. Delaney, N. A. Spaldin and M. Mostovoy, Landau theory of topological defects in multiferroic hexagonal manganites, Nature Materials **13**, 42–49 (2014), https://doi.org/10.1038/NMAT3786

# Supplementary Materials to the manuscript
# "Temperature-Induced Hexagonal-Orthorhombic Phase Transition in Lutetium Ferrite Nanoparticles"

## APPENDIX A. Phase and differential thermal analyses

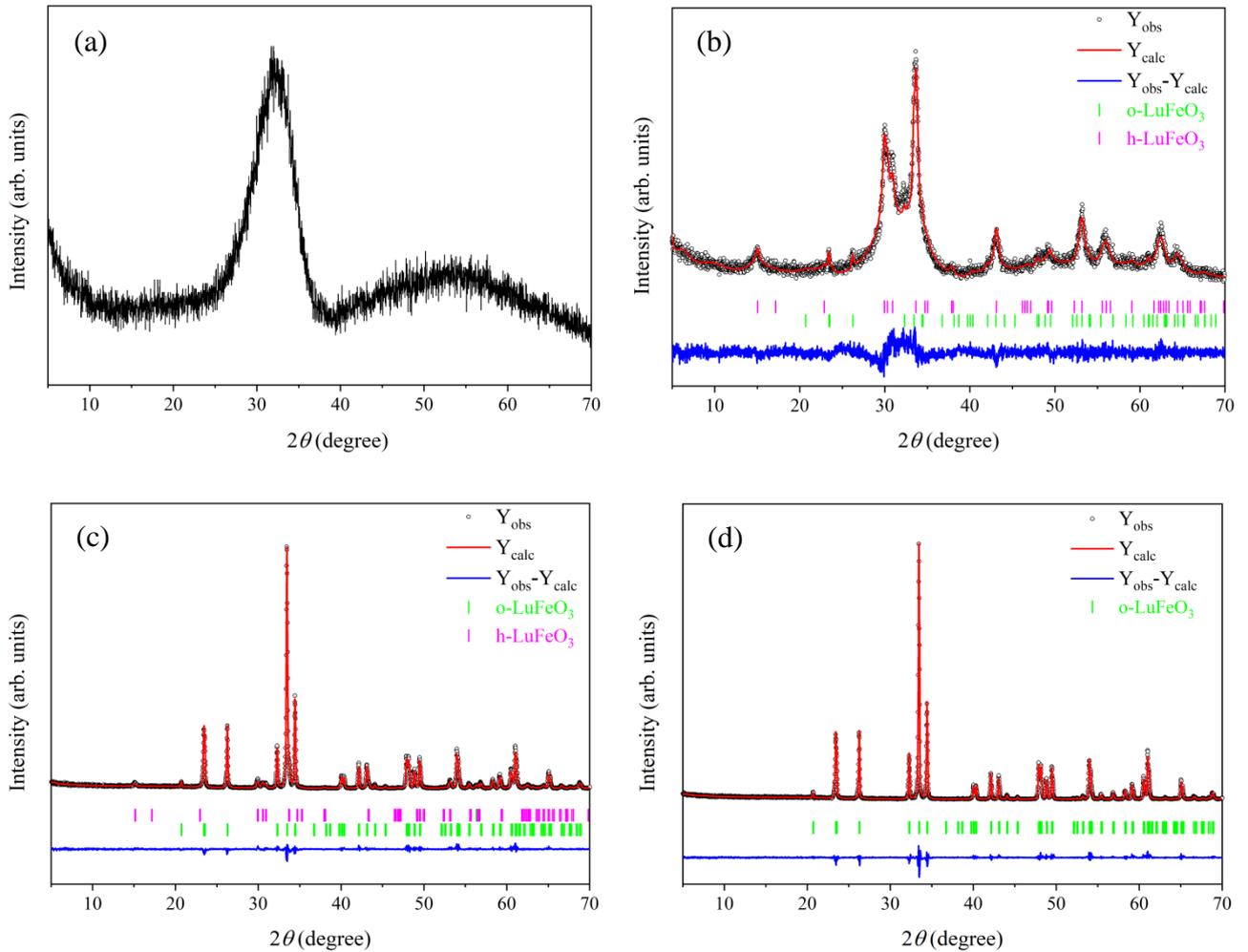

**Fig. A1.** The XRD profile of self-ignited powders calcined at 600°C **(a)** and 700°C **(b)**, 900°C **(c)** and 1100°C **(d)**.

## APPENDIX B. Polar properties of LFO nanoparticles

Below we proposed a phenomenological model of the ferroelectric phase transition in h-LFO, considering the coupling between the structural order parameter trimerization (which is the primary long-range order parameter) and the spontaneous polarization (which is the secondary order parameter). Such approach describes the situation when the structural ordering induces or strongly changes the polar ordering. The approach is widely and successfully used for the description of the sequel and trigger-type phase transitions in ferrodistortive and antiferrodistortive multiferroics, such as (Bi,Sm)FeO$_3$ [1, 2] and (Sr,Ca)TiO$_3$ [3, 4], which perovskite structure in close to the Lu-Fe-O$_6$ structure of the o-LFO.

However, this approach is typically used for bulk materials; and its use for nanoparticles or thin films requires a separate justification. As the justification, we note that the applicability of such approach is determined primarily by the sizes of the nanoparticles, which should significantly exceed a certain characteristic length (namely, the correlation radius of the corresponding order parameter), usually tens of times. For structural phase transitions, including the ferroelectric one, the correlation radius is of the lattice constant order. Thus, our approach is applicable if the nanoparticle dimensions are ten times (or more) larger than the lattice constant. As it follows from the TEM images, shown in **Fig. 3(a)**, and from the sizes of coherent scattering regions, determined by XRD and shown in **Fig. 3(c)**, all these sizes are higher than (10 – 100) nm, which is much higher than 10 lattice constants (~ 5 nm).

It is well-known that a continuous transition between the o-phase and the h-phase is impossible in LFO, because the transition from the Fe-O$_6$ octahedrons to Fe-O$_5$ bipyramids requires a "break" of atomic displacements, instead of being as a set of continuous small atomic displacements (see e.g., Ref. [16]). That is why the energies of o- and h-phases are very different, and they should be considered as allotropic structures. All material below is devoted to the description of the trimerization transition in the h-phase and has no relation to the o-phase.

The background of the theoretical consideration is that the distortion of the high symmetry aristo-phase leading to the transition to the low-symmetry phase could be decomposed into the combination of the aristo-phase eigenmodes (see e.g., Refs. [5] and [6]). Graphical interpretations of the eigen-vectors of different phonon modes are shown in **Fig. B1**. Here the undistorted crystal structure of the h-LFO is plotted using the "stick-and-ball" model.

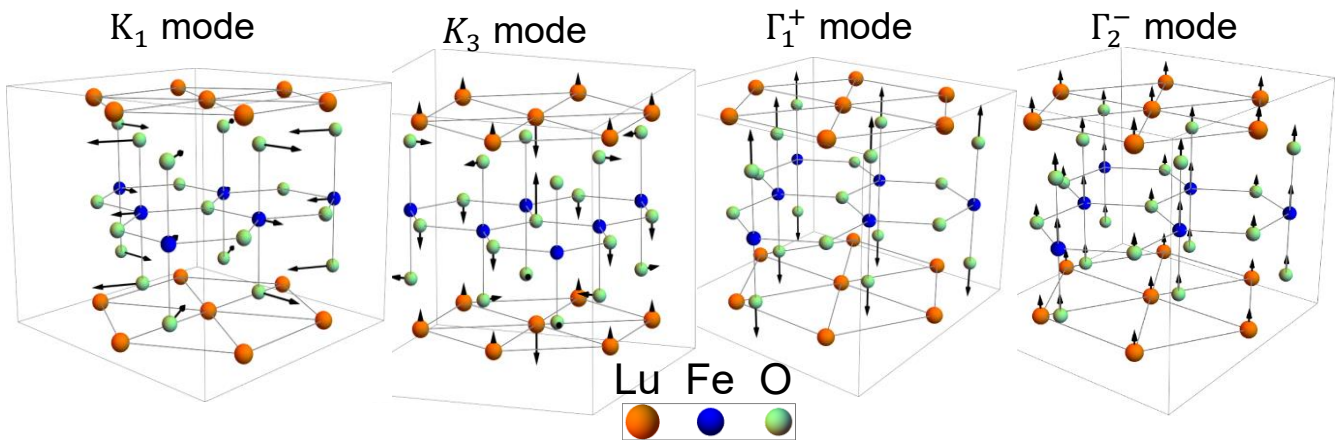

**Figure B1.** The undistorted crystal structure of h-LFO plotted using the "stick-and-ball" model. Arrows represent the displacement vectors of different atomic groups for different phonon modes, denoted near the structures. Adapted from Ref.[7].

Total displacement of atom "$\alpha$" could be decomposed as the mixture of all four modes in the form of the linear combination of corresponding eigen vectors [7]:

$$\vec{U}^{(\alpha)} = Q_{K_3}\vec{U}^{(\alpha)}_{K_3} + Q_{K_1}\vec{U}^{(\alpha)}_{K_1} + Q_{\Gamma_1}\vec{U}^{(\alpha)}_{\Gamma_1} + Q_{\Gamma_2}\vec{U}^{(\alpha)}_{\Gamma_2} \quad \text{(B.1)}$$

Here $\vec{U}^{(\alpha)}_{K_i}$ are the eigen vectors, $Q^{(\alpha)}_{K_i}$ and $Q^{(\alpha)}_{\Gamma_i}$ are the amplitudes of the antipolar and polar phonon modes, superscript $\alpha = \text{Lu, Fe, O}$ and $i = x, y, z$. The modes $K_1$ and $K_3$ are responsible for the trimerization.

Using the decomposition (B.1), we can estimate the spontaneous polarization and rotation angle of h-LFO bi-pyramids. The spontaneous polarization, pointed along Z-axis, can be found from the coordinates of atom "$\alpha$" and their effective charges $q^{(\alpha)}$ on the basis of the effective point-charge model [7]:

$$P_s = \frac{e\,c}{V}\sum_{(\alpha)}\left\{Q^{(\alpha)}_{z\Gamma_2} + Z^{(\alpha)}\right\}q^{(\alpha)}. \quad \text{(B.2)}$$

Here $e$ is the elementary charge, $c$ is the lattice constant, $V$ is the unit cell volume, $q^{(\alpha)}$ are effective charges, $Z^{(\alpha)}$ are the z-coordinates of atom "$\alpha$" in the prototype non-polar phase. The tilt angle of Fe-$O_5$ bi-pyramids can be estimated as follows [7]:

$$\Phi_{K_3} = \arctan\left(\frac{aQ^{(O)}_{xK_3}}{l}\right), \quad \text{(B.3)}$$

where $a$ is the lattice constant, and $l$ is the length of the bond between the Fe-cation and apical oxygen in the prototype non-polar phase.

Taking into account the trigonal symmetry of the system, the Landau-Ginsburg-Devonshire model for the hexagonal ferrite/manganite compounds is

$$\Delta F = a_Q\frac{Q_x^2+Q_y^2}{2} + b_Q\frac{(Q_x^2+Q_y^2)^2}{4} + \left\{c_2 Q_y^2(-3Q_x^2+Q_y^2)^2 + c_1 Q_x^2(Q_x^2-3Q_y^2)^2\right\}\frac{1}{6} + a_P\frac{P^2}{2} + b_P\frac{P^4}{4} - PE +$$
$$\frac{\mu}{2}(Q_x^2+Q_y^2)P^2 - gPQ_y(Q_y^2 - 3Q_x^2) \quad \text{(B.4)}$$

Here $Q_{x,y}$ are components of the structural order parameter $\mathbf{Q}$, responsible for the transition between the 6/mmm and 6mm symmetry phases. $Q_{x,y}$ are the components of the apical oxygen ions displacement (see **Fig. B2**) leading to the trimerization of unit cell of the initial aristo-phase (see **Figs. B2c** and **B2d**). Here we introduced the polarization $P$ and electric field $E$ directed along the hexagonal symmetry axis Z. Below we suppose that the parameters $a_Q < 0$, $c_{1,2} > 0$, $a_P > 0$ and $b_P > 0$.

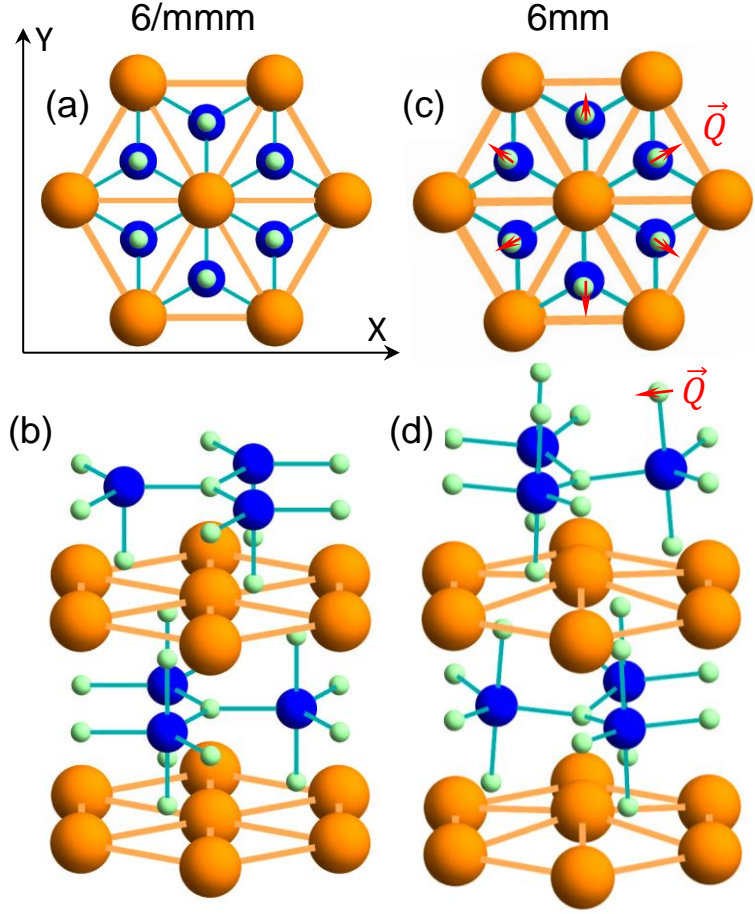

**Figure B2.** "Balls-and-sticks" models of the hexagonal ABO$_3$ in different phases, namely **(a, b)** is the aristo-phase with symmetry 6/mmm; **(c, d)** are the "trimerized" polar phase with 6mm symmetry. Large orange, medium blue and small green spheres represent A cations, B cations and oxygens, respectively. Red arrows show the displacement of the apical oxygens in the bi-pyramids B-O$_5$, which is the primary order parameter of trimerization transition between the phases with the 6/mmm and 6mm symmetries. Models **(a, c)** is the top view along the hexagonal axis, and **(b, d)** is the side view.

Introducing the amplitude and phase of lateral displacement as $Q^2 = Q_x^2 + Q_y^2$, $Q_x = Q \sin(\Phi)$ and $Q_y = Q \cos(\Phi)$, one could rewrite Eq.(B.1) as follows:

$$\Delta F = a_Q \frac{Q^2}{2} + b_Q \frac{Q^4}{4} + [c_1 + (c_2 - c_1)\cos(3\Phi)^2]\frac{Q^6}{6} + a_P \frac{P^2}{2} + b_P \frac{P^4}{4} - PE + \frac{\mu}{2} Q^2 P^2 - gPQ^3\cos(3\Phi)$$
(B.5)

Equations of state are obtained from the minimization of Eq.(B.5) with respect to the amplitude $Q$ of trimerization vector $\boldsymbol{Q}$, its phase $\Phi$ and polarization $P$:

$$a_Q Q + b_Q Q^3 + \{c_1 + (c_2 - c_1)\cos(3\Phi)^2\}Q^5 + \mu Q P^2 - 3gPQ^2\cos(3\Phi) = 0 \quad (B.6a)$$

$$-6\cos(3\Phi)\sin(3\Phi)(c_2 - c_1)\frac{Q^6}{6} + 3gPQ^3\sin(3\Phi) = 0, \quad (B.6b)$$

$$a_P P + b_P P^3 - E + \mu Q^2 P - gQ^3\cos(3\Phi) = 0. \quad (B.6c)$$

In the limit of small polarization, one could solve (B.6c) with respect to polarization:

$$P_z \approx \frac{E+gQ^3\cos(3\Phi)}{a_P+\mu Q^2}. \tag{B.7a}$$

In the absence of the electric field, using Eq.(B.7a) one could exclude the polarization from the Eq (B.6a) and (B.6b):

$$a_Q Q + b_Q Q^3 + \left\{c_1 + \left(c_2 - c_1 - \frac{3g^2}{a_P+\mu Q^2}\right)\cos(3\Phi)^2\right\}Q^5 + \mu Q \left(\frac{gQ^3\cos(3\Phi)}{a_P+\mu Q^2}\right)^2 = 0, \tag{B.7b}$$

$$Q^6 \left\{(c_2 - c_1) - \frac{3g^2}{a_P+\mu Q^2}\right\}\cos(3\Phi)\sin(3\Phi) = 0. \tag{B.7c}$$

It is seen from Eq.(B.7c) that there are twelve extrema, corresponding to the following $\Phi$ values:

$$\Phi = \frac{\pi n}{6}, \quad n = 0 \div 11. \tag{B.8}$$

However, only half of them are stable minima, and the other are the saddle points separating the minima points. At $c_2 - c_1 - \frac{3g^2}{a_P+\mu Q^2} > 0$ the minima positions are

$$\Phi = \frac{\pi}{6} + \frac{\pi n}{3}, \quad n = 0 \div 5. \tag{B.9a}$$

Corresponding polarization and equation for the trimerization parameter are

$$P \approx 0, \tag{B.9b}$$

$$a_Q Q + b_Q Q^3 + c_1 Q^5 = 0. \tag{B.9c}$$

However, for the rather large values of $g$ or $c_2 - c_1 < 0$, the minimal values are

$$\Phi = \frac{\pi n}{3}, \quad n = 0 \div 5, \tag{B.10a}$$

while Eq.(B.9a) corresponds to saddle points. In this case the polarization and the equation for the trimerization parameter are

$$P \approx \frac{E+gQ^3(-1)^n}{a_P+\mu Q^2}, \tag{B.10b}$$

$$a_Q Q + b_Q Q^3 + \left\{c_2 - \frac{3g^2}{a_P+\mu Q^2}\right\}Q^5 + \mu Q \left(\frac{gQ^3}{a_P+\mu Q^2}\right)^2 = 0. \tag{B.10c}$$

The renormalization of expansion coefficients in Eqs.(B.10b) and (B.10c) is due to the surface tension and striction effects:

$$a_P(T) \to a_P(T) + R_P \left(\frac{\gamma_S^{ho}}{D} + \frac{\delta_S^{ho}}{D^2}\right), \tag{B.11a}$$

$$a_Q(T) \to a_Q(T) + R_Q \left(\frac{\gamma_S^{ho}}{D} + \frac{\delta_S^{ho}}{D^2}\right). \tag{B.11b}$$

Here $R_P$ and $R_Q$ are the electrostriction and trimerization/rotation striction coefficients, respectively. We put $|R_P| \leq 0.5$ m$^4$/C$^2$ and $|R_Q| \leq 10^{22}$ m$^{-2}$ according to the structural data. Interface energy parameters $\gamma_S^{ho} = -2.05697 \cdot 10^{-3}$ J/m$^2$, $\delta_S^{ho} = 6.855 \cdot 10^{-12}$ J/m [see Eq.(5b) in the main text].

Since the ferroelectricity of hexagonal ferrites is incipient, coefficient $a_P$ is rather big and positive, while the negative coefficient $a_Q$ should become zero with the temperature decrease, because $a_Q(T) = a_{Q0}(T - T_C)$, where $T_C \cong 1050$ K. Therefore, electrostriction coefficient $R_P$ does not matter much in Eq.(B.11a), while the coupling of the strain with the trimerization order parameter are very important in Eq.(B.11b).

**Table BI.** Values of the free energy expansion coefficient at $T = 0$ taken from Ref.[8].

| $a_Q$ | $b_Q$ | $c_Q$ | $c_\Phi$ | $a_P$ | $g$ | $\mu$ |
|---|---|---|---|---|---|---|
| -2.626 eVÅ$^{-2}$ | 3.375 eVÅ$^{-4}$ | 0.117 eVÅ$^{-6}$ | 0.108 eVÅ$^{-6}$ | 0.866 eVÅ$^{-2}$ | 1.945 eVÅ$^{-4}$ | 9.931 eVÅ$^{-4}$ |

Note that $c_Q \equiv \frac{c_1+c_2}{2}$, $c_\Phi \equiv \frac{c_1-c_2}{2}$